\documentclass[aps,prd,reprint,onecolumn,superscriptaddress,nofootinbib]{revtex4}
\usepackage[T1]{fontenc}
\usepackage[utf8]{inputenc}
\usepackage[english]{babel}
\usepackage{microtype,booktabs,graphicx,amssymb,amstext,amsmath,amsthm,mathtools,braket,verbatim,cancel,epsfig,physics}
\usepackage[figurename=FIG.]{caption}
\DeclareCaptionLabelSeparator{period}{.\space}
\usepackage[english]{varioref}
\usepackage[hidelinks]{hyperref}
\DeclareGraphicsExtensions{.eps}
\usepackage[output-decimal-marker={,}]{siunitx}
\usepackage[autostyle,italian=guillemets]{csquotes}
\newcommand{\IDG}{Infinite Derivative Gravity}
\newcommand{\GR}{General Relativity}
\newcommand{\SMoC}{Standard Model of Cosmology}
\newcommand{\FLRW}{Friedmann--Lema\^{\i}tre--Robertson--Walker}
\DeclareMathOperator{\diag}{diag}

\begin{document}
\title{Can nonlocal gravity  really explain dark energy?}

\author{Salvatore Capozziello}
\email{capozziello@unina.it}
\affiliation{Dipartimento di Fisica ``E. Pancini'', Universit\`a di Napoli ``Federico II'', Via Cinthia 9, 80126 Napoli, Italy.}
\affiliation{Scuola Superiore Meridionale, Largo San Marcellino, 10, 80138, Napoli, Italy}
\affiliation{INFN Sezione di Napoli, Complesso Universitario di Monte Sant'Angelo, Edificio 6, Via Cinthia, 80126, Napoli, Italy.}

\author{Anupam Mazumdar}
\email{anupam.mazumdar@rug.nl}
\affiliation{Van Swinderen Institute, University of Groningen, 9747 AG Groningen, The Netherlands.}

\author{Giuseppe Meluccio}
\email{giuseppe.meluccio-ssm@unina.it}
\affiliation{Scuola Superiore Meridionale, Largo San Marcellino, 10, 80138, Napoli, Italy}
\affiliation{INFN Sezione di Napoli, Complesso Universitario di Monte Sant'Angelo, Edificio 6, Via Cinthia, 80126, Napoli, Italy.}

\date{\today}

\begin{abstract}
	In view to scrutinize the idea that nonlocal modifications of General Relativity could dynamically address the  dark energy problem, we investigate the evolution of the Universe at  infrared scales  as an Infinite Derivative Gravity model of the Ricci scalar, without introducing the cosmological constant $\Lambda$ or any scalar field. The accelerated expansion of the late Universe is shown to be compatible with the emergence of nonlocal gravitational effects at sufficiently low energies. A technique for circumventing the mathematical complexity of the nonlocal cosmological equations is developed and, after drawing a connection with the Starobinsky gravity, verifiable predictions are considered, like a possible decreasing in the strength of the effective gravitational constant. In conclusion, the emergence of nonlocal gravity corrections at given scales could be an efficient mechanism to address the dark energy problem.
\end{abstract}

\maketitle

\def\sectionautorefname{Appendix}

\section{Introduction}
There are many experimental and observational evidences that \GR, assumed as the fundamental theory of gravity,  is valid at several energy scales. However, despite the enormous success of the Einstein picture, it lacks in describing the whole cosmic history when stretched to ultraviolet (UV) and infrared (IR) regimes. These shortcomings can be related to the emergence of tensions in cosmological parameters \cite{Abdalla}.

The issues for the validity of \GR~at UV scales are well known: in fact, the Quantum Field Theory applied to gravity yields incurable divergences beyond the second loop order of renormalization \cite{goroff:ultraviolet}, implying the impossibility of a self-consistent quantum description of the gravitational interaction. Nevertheless, the interest to describe \GR~at UV scales has not waned, in spite of the fact that a fully-fledged theory of Quantum Gravity is not available yet. From a cosmological point of view, the most active areas of research in this direction include Cosmic Inflation  \cite{Starobinsky,linde:inflationary,guth:inflationary,linde:new,linde:chaotic} and Quantum Cosmology \cite{hartle:wave,vilenkin:quantum,halliwell:introductory, vanPutten1}.

At astrophysical scales, the predictions of \GR~are greatly successful in giving accurate descriptions of compact objects, like neutron stars and black holes, addressing solar system dynamics, and also the cosmic evolution of radiation- and matter-dominated eras. However, also at the IR regime \GR~suffers some severe shortcomings, which lead to the introduction of dark matter and dark energy that, up to now, have no definitive counterparts at fundamental particle level. The latter is the appellative of some unknown source of energy driving the accelerated expansion of the Universe in the current cosmological era. In order to describe the observed rate of expansion, it seems that such energy density -- which comprises almost 70\% of the energy content of the Universe \cite{planck:results} -- should possess two main features: it should be constant and it should exert a negative pressure. Since, to date, these features cannot be ascribed to any known (and detected) particle, the easiest way to embed a description of dark energy into the \SMoC~is to resort to the addition of a cosmological constant $\Lambda$ into the Einstein field equations of \GR.

Despite appearing a very simple explanation, the introduction of the cosmological constant term raises a few important questions. First of all, there is an issue with understanding the physical source of this energy as well as its observed value; in particular, attempts to link it to the energy density of  vacuum produce theoretical expectations  which exceed observational limits by more or less  120 orders of magnitude \cite{weinberg:cosmological}. Secondly, it might be the case that \GR~should be, in some way, ``extended'' for accounting of cosmological dynamics at IR and UV scales \cite{weinberg:cosmological,
conroy:generalised, vanPutten2, extended,oikonomou, Clifton,SD,Rept, Sotiriou, Vas}.

For the above reasons, last decades have seen a surge of interest in modifications of \GR~in view to change and improve the IR behavior, while retaining successes achieved at other scales \cite{capozziello:overview}. Among these approaches, with the goal of improving the gravitational interaction with quantum corrections, nonlocal theories of gravity represent a class of theories where various nonlocal operators are added to the Einstein--Hilbert action assuming that such corrections arise naturally as quantum loop effects. More specifically, \IDG~theories adopt transcendental functions of the covariant d'Alembert operator $\Box$ (or  its inverse  $\Box^{-1}$), thus introducing \emph{infinite} derivative or integral operators respectively into dynamics. The former family of theories  provides regular black hole and Big Bang solutions \cite{modesto:super,modesto:super-quantum,briscese:inflation,biswas:bouncing,modesto:black,biswas:stable,Novikov1,Novikov2,Novikov3} as well as renormalizable and unitary Quantum Gravity models \cite{Kuzmin, Tomboulis, Spyridon, Lambiase1,Lambiase2}, while the latter set of theories has been conceived mainly for an alternative description of the accelerated cosmic expansion, possibly shedding light on a deeper understanding of the cosmological constant problem \cite{deser:nonlocal1,deser:nonlocal2,arkani:non-local,barvinsky:nonlocal,nojiri:screening,rocco,odintsov1,odintsov2,odintsov3, Lev}. The most general action for  ghost-free theories of gravity  has  been considered in Ref. \cite{biswas:towards}. Nonlocal massive gravity has been considered in \cite{ModestoTsu,Jaccardgla}. 

An important test for such theories could be selecting further polarizations in gravitational waves as a signature of nonlocal terms \cite{capriolo1,capriolo2,maggioreGW} or characteristic lengths in the large scale structure. Specifically, the general question that nonlocality  could supply dark matter  has been recently discussed in Ref. \cite{Wood}. In particular,  the fact that  nonlocal terms can be related to effective lengths  and masses could be the dynamical mechanism giving  rise to dark matter  at astrophysical  scales as reported in \cite{kostas} for our Galaxy and in \cite{filippo1,filippo2} for galaxy clusters.

Specific models of IR modifications of \GR~have recently been proposed \cite{maggiore:phantom,foffa:cosmological,maggiore:nonlocal,jaccard:nonlocal}. In their approach, an IR mass scale parameter is introduced by means of nonlocal terms in the action retaining general covariance. Even though initially this approach was considered as an attempt to construct a consistent Quantum Field Theory of massive gravity, the modified Einstein field equations can be understood as effective equations of motion  emerging from some more fundamental dynamics.

In this paper, we follow a similar approach trying to build up a model that generates a dynamical mechanism for dark energy, without introducing the cosmological constant or any scalar field. We look for IR cosmological solutions to the full equations of motion derived from the nonlocal action presented in \cite{conroy:generalised} and restricted to the quadratic terms in the Ricci scalar $R$, i.e. without truncating the action at a given order of the IR mass scale parameter \cite{vanPutten2}. In doing so, we explore the effects of IR modifications of \GR~stemming from \IDG~as indicated by an action containing an infinite number of  nonlocal operators $\Box^{-1}$, laying emphasis on the cosmic expansion of the late Universe.

The paper is organised as follows. In Sec. \ref{sec:nonlocal}, we define the nonlocal gravitational action and derive its corresponding equations of motion. In Sec. \ref{sec:ansatz}, we outline a way to simplify the cosmological equations and present an approximate IR solution. Finally, we discuss the cosmological solution in Sec. \ref{sec:dark-energy} and investigate whether it is possible to produce testable predictions of late accelerated expansion (i.e. dark energy) in terms of nonlocal effects. We draw conclusions and outline perspectives in Sec. \ref{sec:conclusions}. Details of calculations are reported in \autoref{sec:appendix}.

The parameters of the \SMoC~are taken from \cite{planck:results}. The adopted metric signature is $(-,+,+,+)$. We assume $c=1$ along the draft.

\section{Nonlocal Gravity Cosmology}
\label{sec:nonlocal}
\subsection{The IR nonlocal action}
Considering terms involving the Ricci scalar only, the effective nonlocal gravitational action of \IDG~can be expressed as \cite{conroy:generalised}
\begin{equation}
	\label{eq:action}
	S=\int d^{4}x\,\frac{\sqrt{-g}}{2}[M_\textup{P}^2R+R\bar{\mathcal{F}}_1(\Box)R],
\end{equation}
where
\begin{equation}
	\bar{\mathcal{F}}_1(\Box)=\sum_{n=1}^\infty f_{1_{-n}}\Box^{-n},\qquad f_{1_{-n}}\equiv\tilde{f}_{1_{-n}}M^{2n},
\end{equation}
where $\tilde{f}_{1_{-n}}$ are dimensionless constants; $M_\textup{P}=2.44\cdot10^{18}$\,GeV is the reduced Planck mass and $M$ is an IR mass scale. In the UV limit, $M$ can be neglected with respect to $M_\textup{P}$ and the action reduces to the Einstein--Hilbert action of \GR.

The inverse d'Alembert operator $\Box^{-1}$ can be expressed in terms of its Green function as
\begin{equation}
	(\Box^{-1}j)(x)=f_\textup{hom}(x)+\int d^{4}y\,\sqrt{-g(y)}G(x,y)j(y),\qquad \Box_xG(x,y)=\frac{1}{\sqrt{-g(x)}}\delta^4(x-y),
\end{equation}
where $f_\textup{hom}$ is some homogeneous solution, i.e. any solution satisfying $\Box f_\textup{hom}(x)=0$. The (retarded) Green function of the Minkowski spacetime is given by
\begin{equation}
	G(x,y)=G(x-y)=-\frac{1}{4\pi\abs{\vec{x}-\vec{y}}}\delta(x^0-y^0-\abs{\vec{x}-\vec{y}})\,.
\end{equation}
A solution of the equation $\Box f(x)=j(x)$, with a vanishing homogeneous solution, is thus\footnote{The condition $f_\textup{hom}(x)=0$ is needed in order to achieve a generally covariant solution \cite{conroy:generalised}.}
\begin{equation}
	f(x)=-\frac{1}{4\pi}\int d^{4}y\,\sqrt{-g(y)}\frac{j(y)}{\abs{\vec{x}-\vec{y}}}\delta(x^0-y^0-\abs{\vec{x}-\vec{y}}).
\end{equation}
The field equations obtained through the variational principle are \cite{conroy:generalised}
\begin{equation}
	\label{eq:eq-motion}
	T_{\mu\nu}=M_\textup{P}^2G_{\mu\nu}+2G_{\mu\nu}\bar{\mathcal{F}}_1(\Box)R+\frac{1}{2}g_{\mu\nu}R\bar{\mathcal{F}}_1(\Box)R-2(\nabla_\mu\nabla_\nu-g_{\mu\nu}\Box)\bar{\mathcal{F}}_1(\Box)R+\Theta_{\mu\nu}^1-\frac{1}{2}g_{\mu\nu}\bigl(\Theta_\sigma^{1\sigma}+\bar{\Theta}^1\bigr),
\end{equation}
with
\begin{subequations}
	\begin{align}
		\Theta_{\mu\nu}^1&\equiv\sum_{n=1}^\infty f_{1_{-n}}\sum_{l=0}^{n-1}\nabla_\nu\Box^{-l-1}R\nabla_\mu\Box^{-n+l}R,\\
		\bar{\Theta}^1&\equiv\sum_{n=1}^\infty f_{1_{-n}}\sum_{l=0}^{n-1}\Box^{-l-1}R\Box^{-n+l+1}R.
	\end{align}
\end{subequations}
 Here  $G_{\mu\nu}=R_{\mu\nu}-\frac{1}{2}g_{\mu\nu}R$ is the Einstein tensor. 
 The trace of the field equations is
\begin{equation}
	\label{eq:trace-eq}
	-M_\textup{P}^2R+6\Box \bar{\mathcal{F}}_1(\Box)R-\Theta_\sigma^{1\sigma}-2\bar{\Theta}^1=T.
\end{equation}
The nonlocal action can also be recast in terms of $n$ additional scalar fields $\phi_n$ with the definition
\begin{equation}
	\phi_n\equiv\Box^{-n}R,
\end{equation}
from which follows that
\begin{equation}
	R\bar{\mathcal{F}}_1(\Box)R=R\sum_{n=1}^\infty f_{1_{-n}}\Box^{-n}R=R\sum_{n=1}^\infty f_{1_{-n}}\phi_n;
\end{equation}
therefore, this model can be considered as equivalent to a generalised scalar-tensor theory with $n$ classical scalar fields non-minimally coupled to gravity.

\subsection{Nonlocality in a homogeneous, isotropic and flat Universe}
The \FLRW~metric describing a homogeneous, isotropic and spatially flat Universe  is given by
\begin{equation}
	\label{eq:metric}
	g_{\mu\nu}=\diag\Bigl(-1,a^2(t),a^2(t),a^2(t)\Bigr),
\end{equation}
where $a(t)$ is the scale factor. The expression of the corresponding Ricci scalar is
\begin{equation}
	\label{eq:ricciscalar}
	R=6\frac{a\ddot{a}+\dot{a}^2}{a^2}.
\end{equation}
In this metric, when acting upon a function of time \emph{only}, $f(t)$, the d'Alembert operator and its (retarded) inverse are given respectively by \cite{deser:nonlocal1}
\begin{align}
	(\Box f)(t)&=-a^{-3}\partial_0[a^3\partial_0f(t)],\\
	(\Box^{-1}f)(t)&=-\int_0^t dt'\,a^{-3}(t')\int_0^{t'} dt''\,a^3(t'')f(t'').
\end{align}
Therefore, the expression
\begin{equation}
	\label{eq:box-1R}
	(\Box^{-1}R)(x^0)=-\int_0^{x^0} dt'\,a^{-3}(t')\int_0^{t'} dt''\,a^3(t'')R(t''),
\end{equation}
which is given by a double \emph{time} integral, can be contrasted with the one deriving from the general definition of the inverse of the d'Alembert operator in the Minkowski spacetime,
\begin{equation}
	(\Box^{-1}R)(x^0)=-\frac{1}{4\pi}\int d^{4}y\,a^3(y^0)\frac{R(y^0)}{\abs{\vec{x}-\vec{y}}}\delta(x^0-y^0-\abs{\vec{x}-\vec{y}})=-\frac{1}{4\pi}\int d^{3}y\,a^3(x^0-\abs{\vec{x}-\vec{y}})\frac{R(x^0-\abs{\vec{x}-\vec{y}})}{\abs{\vec{x}-\vec{y}}},
\end{equation}
which, on the contrary, is given by a triple spatial integral; this insight suggests that, in a cosmological context, nonlocality can be conceived as the effect of interactions that happen in the same place at different cosmic times, other than in different positions in space at a given cosmic time.

The cosmological equations follow from the field equations \eqref{eq:eq-motion} applied to the flat \FLRW~metric \eqref{eq:metric}. Because of the symmetries of this metric, there exist only two independent cosmological equations, namely the `00'-component and the `11'-component of \eqref{eq:eq-motion}, which, respectively, describe the time evolution of the energy density $\rho$ and pressure $p$ of a perfect fluid. These two equations can then be combined by the equation of state for a perfect fluid, $p=w\rho$. Another cosmological equation is given by the trace of the field equations; it can be combined with one of these two. A detailed derivation of such integro-differential cosmological equations, i.e. the Eqs. \eqref{eq:cosmo1} and \eqref{eq:cosmo2}, is provided in  \autoref{sec:appendix}.

\section{The formal localisation}
\label{sec:ansatz}
\subsection{Ansatz for the nonlocal operators}
Following Refs. \cite{koshelev:bouncing,biswas:bouncing}, a formal integration of equations of motion can be achieved by imposing
\begin{equation}
	\label{eq:ansatz}
	\Box^{-1}R=r_1R+r_2,\qquad r_1\ne0,
\end{equation}
which allows to recast the nonlocal operators in a local form, where the information about nonlocality is contained in the parameters $r_1$ and $r_2$. This implies that
\begin{align}
	\label{eq:box-nR}
	\Box^{-n}R&=r_1^n\biggl(R+\frac{r_2}{r_1}\biggr),\qquad n>0,\\
	\label{eq:F_1}
	\bar{\mathcal{F}}_1(\Box)R=\sum_{n=1}^\infty f_{1_{-n}}r_1^n\biggl(R+\frac{r_2}{r_1}\biggr)&=\bar{\mathcal{F}}_1(r_1^{-1})\biggl(R+\frac{r_2}{r_1}\biggr),\qquad \bar{\mathcal{F}}_1(r_1^{-1})=\sum_{n=1}^\infty f_{1_{-n}}r_1^n,
\end{align}
so that Eqs. \eqref{eq:eq-motion} become
\begin{equation}
	\label{eq:eq-motion-ansatz}
	T_{\mu\nu}=M_\textup{P}^2G_{\mu\nu}+2\bar{\mathcal{F}}_1(r_1^{-1})R_{\mu\nu}\biggl(R+\frac{r_2}{r_1}\biggr)-\frac{1}{2}\bar{\mathcal{F}}_1(r_1^{-1})g_{\mu\nu}R\biggl(R+\frac{r_2}{r_1}\biggr)-2\bar{\mathcal{F}}_1(r_1^{-1})(\nabla_\mu\nabla_\nu-g_{\mu\nu}\Box)R+\Theta_{\mu\nu}^1-\frac{1}{2}g_{\mu\nu}\bigl(\Theta_\sigma^{1\sigma}+\bar{\Theta}^1\bigr),
\end{equation}
with (see Eqs. \eqref{eq:theta1}-\eqref{eq:theta2} in \autoref{sec:appendix})
\begin{align}
	\Theta_{\mu\nu}^1&=\delta_\mu^0\delta_\nu^0\dot{R}^2\sum_{n=1}^\infty nf_{1_{-n}}r_1^{n+1},\\
	\Theta_\sigma^{1\sigma}&=-\dot{R}^2\sum_{n=1}^\infty nf_{1_{-n}}r_1^{n+1},\\
	\bar{\Theta}^1&=\biggl(R+\frac{r_2}{r_1}\biggr)^2\sum_{n=1}^\infty nf_{1_{-n}}r_1^n-\frac{r_2}{r_1}\bar{\mathcal{F}}_1(r_1^{-1})\biggl(R+\frac{r_2}{r_1}\biggr).
\end{align}
When applied to the trace Eq. \eqref{eq:trace-eq}, the same ansatz yields
\begin{equation}
	\label{eq:trace-eq-ansatz}
	-M_\textup{P}^2R+6\bar{\mathcal{F}}_1(r_1^{-1})\Box R-\Theta_\sigma^{1\sigma}-2\bar{\Theta}^1=T.
\end{equation}

 Eqs. \eqref{eq:eq-motion-ansatz} and \eqref{eq:trace-eq-ansatz} allow to simplify the nonlocal cosmological Eqs. \eqref{eq:cosmo1} and \eqref{eq:cosmo2} respectively as
\begin{equation}
	\label{eq:cosmo1-ansatz}
	\begin{split}
		&M_\textup{P}^2\biggl[2\frac{\ddot{a}}{a}+(3w+1)\frac{\dot{a}^2}{a^2}\biggr]-(3w-1)\bar{\mathcal{F}}_1(r_1^{-1})\frac{a\ddot{a}-\dot{a}^2}{a^2}\biggl(6\frac{a\ddot{a}+\dot{a}^2}{a^2}+\frac{r_2}{r_1}\biggr)+12\bar{\mathcal{F}}_1(r_1^{-1})\frac{a\ddot{a}^2-5\dot{a}^2\ddot{a}+3a\dot{a}\dddot{a}+a^2\ddddot{a}}{a^3}\\
		&-\frac{w+1}{2}\frac{r_2}{r_1}\bar{\mathcal{F}}_1(r_1^{-1})\biggl(6\frac{a\ddot{a}+\dot{a}^2}{a^2}+\frac{r_2}{r_1}\biggr)+\Biggl[18(w-1)r_1\frac{(a^2\dddot{a}+a\dot{a}\ddot{a}-2\dot{a}^3)^2}{a^6}+\frac{w+1}{2}\biggl(6\frac{a\ddot{a}+\dot{a}^2}{a^2}+\frac{r_2}{r_1}\biggr)^2\Biggr]\sum_{n=1}^\infty nf_{1_{-n}}r_1^n=0
	\end{split}
\end{equation}
and
\begin{equation}
	\label{eq:cosmo2-ansatz}
	\begin{split}
		&M_\textup{P}^2\frac{a\ddot{a}+\dot{a}^2}{a^2}+6\bar{\mathcal{F}}_1(r_1^{-1})\frac{a\ddot{a}^2-5\dot{a}^2\ddot{a}+3a\dot{a}\dddot{a}+a^2\ddddot{a}}{a^3}-\frac{r_2}{3r_1}\bar{\mathcal{F}}_1(r_1^{-1})\biggl(6\frac{a\ddot{a}+\dot{a}^2}{a^2}+\frac{r_2}{r_1}\biggr)\\
		&-\biggl[6r_1\frac{(a^2\dddot{a}+a\dot{a}\ddot{a}-2\dot{a}^3)^2}{a^6}-\frac{1}{3}\biggl(6\frac{a\ddot{a}+\dot{a}^2}{a^2}+\frac{r_2}{r_1}\biggr)^2\Biggr]\sum_{n=1}^\infty nf_{1_{-n}}r_1^n=-\frac{T}{6};
	\end{split}
\end{equation}
clearly, no integral operator appears anymore, but still these cosmological equations appear very hard to solve, as they are fourth order nonlinear differential equations for the scale factor.

In the matter-dominated era, we have $w=0$, $T=-\rho_M$ and $\rho_M=\Omega_m\rho_ca_0^3a^{-3}$, where $\Omega_m=0.315$ is the density parameter of nonrelativistic matter at the present cosmic time and $\rho_c=3.66\cdot10^{-47}$\,GeV\textsuperscript{4} is the critical density of the Universe, so cosmological Eqs. \eqref{eq:cosmo1-ansatz} and \eqref{eq:cosmo2-ansatz} become respectively
\begin{equation}
	\label{eq:cosmo1-matter}
	\begin{split}
		&M_\textup{P}^2\biggl(2\frac{\ddot{a}}{a}+\frac{\dot{a}^2}{a^2}\biggr)+\bar{\mathcal{F}}_1(r_1^{-1})\frac{a\ddot{a}-\dot{a}^2}{a^2}\biggl(6\frac{a\ddot{a}+\dot{a}^2}{a^2}+\frac{r_2}{r_1}\biggr)+12\bar{\mathcal{F}}_1(r_1^{-1})\frac{a\ddot{a}^2-5\dot{a}^2\ddot{a}+3a\dot{a}\dddot{a}+a^2\ddddot{a}}{a^3}\\
		&-\frac{r_2}{2r_1}\bar{\mathcal{F}}_1(r_1^{-1})\biggl(6\frac{a\ddot{a}+\dot{a}^2}{a^2}+\frac{r_2}{r_1}\biggr)-\Biggl[18r_1\frac{(a^2\dddot{a}+a\dot{a}\ddot{a}-2\dot{a}^3)^2}{a^6}-\frac{1}{2}\biggl(6\frac{a\ddot{a}+\dot{a}^2}{a^2}+\frac{r_2}{r_1}\biggr)^2\Biggr]\sum_{n=1}^\infty nf_{1_{-n}}r_1^n=0
	\end{split}
\end{equation}
and
\begin{equation}
	\label{eq:cosmo2-matter}
	\begin{split}
		&M_\textup{P}^2\frac{a\ddot{a}+\dot{a}^2}{a^2}+6\bar{\mathcal{F}}_1(r_1^{-1})\frac{a\ddot{a}^2-5\dot{a}^2\ddot{a}+3a\dot{a}\dddot{a}+a^2\ddddot{a}}{a^3}-\frac{r_2}{3r_1}\bar{\mathcal{F}}_1(r_1^{-1})\biggl(6\frac{a\ddot{a}+\dot{a}^2}{a^2}+\frac{r_2}{r_1}\biggr)\\
		&-\biggl[6r_1\frac{(a^2\dddot{a}+a\dot{a}\ddot{a}-2\dot{a}^3)^2}{a^6}-\frac{1}{3}\biggl(6\frac{a\ddot{a}+\dot{a}^2}{a^2}+\frac{r_2}{r_1}\biggr)^2\Biggr]\sum_{n=1}^\infty nf_{1_{-n}}r_1^n=\frac{1}{6}\Omega_m\rho_ca_0^3a^{-3}.
	\end{split}
\end{equation}

It is worth noticing that, if the Ricci scalar expression is implicitly assumed into the trace Eq. \eqref{eq:trace-eq-ansatz}, such an equation can be recast as a Klein--Gordon equation for $R$, where the source term is improved by nonlocal curvature contributions as well:
\begin{equation}
	\bigl[6\bar{\mathcal{F}}_1(r_1^{-1})\Box-M_\textup{P}^2\bigr]R=T-\frac{2r_2}{r_1}\bar{\mathcal{F}}_1(r_1^{-1})\biggl(R+\frac{r_2}{r_1}\biggr)-\Biggl[r_1\dot{R}^2-2\biggl(R+\frac{r_2}{r_1}\biggr)^2\Biggr]\sum_{n=1}^\infty nf_{1_{-n}}r_1^n.
\end{equation}
An important remark is in order here. 

\subsection{An approximate IR solution}
Even though the ansatz for the nonlocal operators allows to considerably simplify the equations of motion, it partially alleviates the complexity of the model and, in particular, the peculiarity of nonlocal effects. In fact, not only the nonlocality contributions are parameterized in $r_1$ and $r_2$, but also the mathematical complexity is shifted to the ansatz itself: any possible solution of Eqs. \eqref{eq:cosmo1-ansatz} and \eqref{eq:cosmo2-ansatz} must first satisfy the relation \eqref{eq:ansatz}  for some values of $r_1$ and $r_2$. Indeed, the explicit form of the ansatz, in terms of the scale factor $a(t)$, must satisfy a nonlinear integro-differential equation containing derivatives up to second order as well as a double time integral:
\begin{equation}
	-6\int_0^t dt'\,a^{-3}(t')\int_0^{t'} dt''\,(a^2\ddot{a}+a\dot{a}^2)(t'')=6r_1\frac{a\ddot{a}+\dot{a}^2}{a^2}+r_2,\qquad r_1\ne0\,.
\end{equation}
Also this form is quite difficult to be reduced to simple analytic solutions, hence we will search for approximate solutions still retaining all the information on nonlocal effects from \IDG. For example, the following form of the scale factor can be adopted to implement the technique:
\begin{equation}
	\label{eq:a}
	a(t)=\bar{a}e^{\frac{\lambda}{2}t^2}.
\end{equation}
The corresponding Hubble parameter is
\begin{equation}
	\label{eq:H}
	H(t)\equiv\frac{\dot{a}}{a}=\lambda t.
\end{equation}
For $\lambda>0$ this solution represents a possible way to account for the accelerated expansion of late Universe and so, in the IR limit, this toy model can dynamically describe dark energy.

The free parameters of the model are the IR mass scale $M$, the parameter $\lambda$ of the scale factor and the numerical coefficients $\tilde{f}_{1_{-n}}$. The IR mass has to be $M\sim\sqrt\Lambda\sim H_0$ for consistency with observations, where $\Lambda=4.24\cdot10^{-84}$\,GeV\textsuperscript{2} is the cosmological constant and $H_0\approx1.44\cdot10^{-42}$\,GeV is the Hubble constant.  Since $\Lambda\sim10^{-84}$\,GeV\textsuperscript{2}, this means that we can assume $M\sim10^{-42}$\,GeV.

\begin{figure}
	\begin{center}
		\includegraphics[scale=0.55]{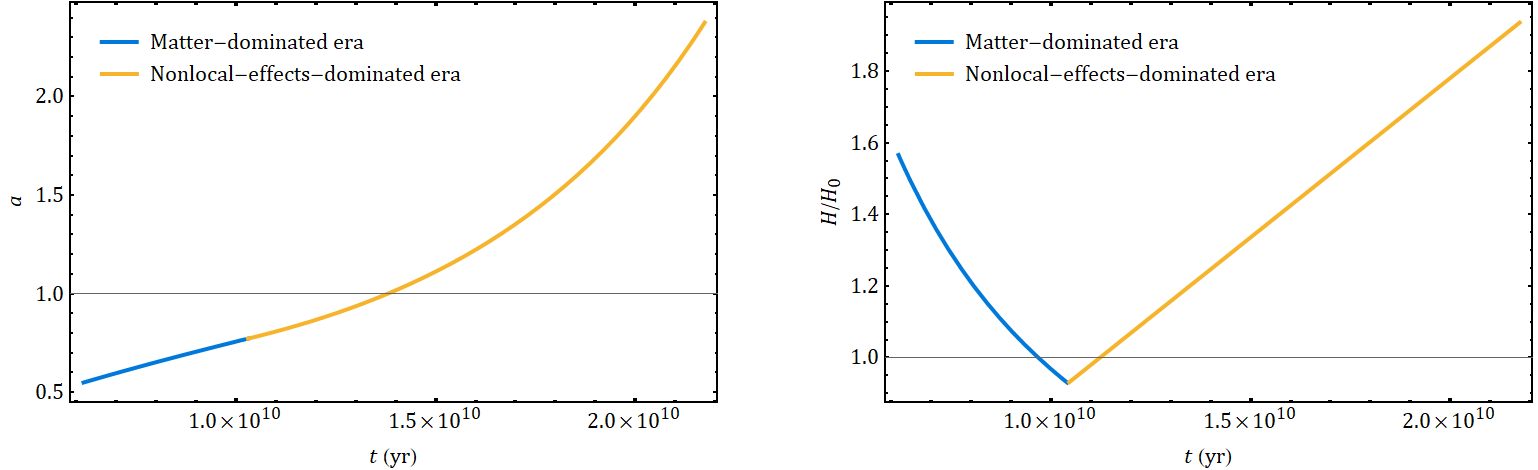}
		\caption{The evolution of the scale factor (on the left) and   the Hubble parameter normalized with respect to the Hubble constant (on the right) in the late Universe.\label{fig:a-H}}
	\end{center}
\end{figure}

In the \SMoC, the transition from the matter-dominated to the dark-energy-dominated era happens at
\begin{equation}
	z_*=0.295,\qquad a_*=0.772,\qquad t_*=10.3\,\text{Gyr}.
\end{equation}
If the model we are considering  has to describe dynamically dark energy, then its corrections to Einstein’s \GR~should be negligible until $t_*$ -- so that all predictions of the \SMoC~remain unaltered -- and, from then on, its effects should become observable, possibly resulting in predictions  discriminating from those of the \SMoC. Since the model does not introduce new matter degrees of freedom, nonrelativistic matter contributes to the dynamics also for $t>t_*$, even though its effects are small when compared to the nonlocal  effects: in this case, we can refer to a nonlocal-effects-dominated era. See the plots of Eqs. \eqref{eq:a} and \eqref{eq:H} in FIG. \ref{fig:a-H}.
The parameter $\lambda$ can thus be fixed by requiring that the scale factor for the matter-dominated era ($a(t)\propto t^{2/3}$) and that for the subsequent accelerated expansion match at $t_*$. First of all, by imposing the boundary condition $a(t_0)=a_0$, where $t_0=13.8$\,Gyr is the current age of the Universe, one finds the expression
\begin{equation}
	\label{eq:scalefactor}
	a(t)=a_0e^{\frac{\lambda}{2}(t^2-t_0^2)};
\end{equation}
then, the matching condition gives
\begin{equation}
	a(t_*)=a_0e^{\frac{\lambda}{2}(t_*^2-t_0^2)}=a_*\Rightarrow\lambda=\frac{2\ln\bigl(\frac{a_0}{a_*}\bigr)}{t_0^2-t_*^2},
\end{equation}
that is $\lambda=6.14\cdot10^{-3}$\,Gyr$^{-2}=2.67\cdot10^{-84}$\,GeV\textsuperscript{2}. Therefore, while $M_\textup{P}^2\sim10^{37}$\,GeV\textsuperscript{2} is the UV mass scale, $M^2$ and $\lambda$ are both IR mass parameters with $M^2\sim\lambda\sim10^{-84}$\,GeV\textsuperscript{2}.

\begin{figure}
	\begin{center}
		\includegraphics[scale=0.4]{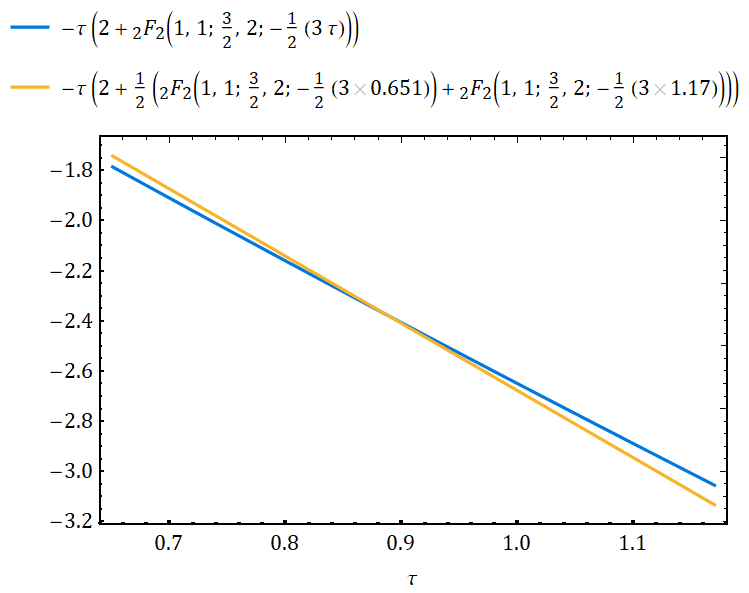}
		\caption{Comparison of $(\Box^{-1}R)(\tau)$ and its approximation in the time interval $\tau_*\leq\tau\leq\tau_0$.\label{fig:2F2}}
	\end{center}
\end{figure}

It is convenient to introduce the dimensionless time parameter $\tau\equiv\lambda t^2$, thereby $\tau(t_*)\equiv\tau_*=0.651$ and $\tau(t_0)\equiv\tau_0=1.17$. One finds
\begin{align}
	\label{eq:Ransatz}
	R(\tau)&=12\lambda\tau+6\lambda,\\
	\label{eq:boxRansatz}
	(\Box R)(\tau)&=-72\lambda^2\tau-24\lambda^2,\\
	\label{eq:box-1Ransatz}
	(\Box^{-1}R)(\tau)&=-\tau\biggl[2+{}_2F_2\biggl(1,1;\frac{3}{2},2;-\frac{3}{2}\tau\biggr)\biggr],
\end{align}
where ${}_2F_2$ is a generalised hypergeometric function. This scale factor does not satisfy exactly  the ansatz \eqref{eq:ansatz}. However, in the interval $\tau_*\leq\tau\leq\tau_0$, the function ${}_2F_2$ is  slowly-varying in time and thus approximates with its mean value $\approx2/3$. See FIG. \ref{fig:2F2}. It is worth noticing that the relative errors are less than 3\%.

As a result, the values of the parameters for which the ansatz \eqref{eq:ansatz} is approximately verified are
\begin{equation}
	\label{eq:r1-r2}
	r_1\approx-\frac{2}{9\lambda}\equiv\frac{1}{\bar{\lambda}},\qquad r_2\approx\frac{4}{3}.
\end{equation}
From  Eqs. \eqref{eq:box-nR} and \eqref{eq:F_1}, we see that this result implies
\begin{align}
	\Box^{-n}R&=\bar{\lambda}^{-n}(R-6\lambda),\\
	\label{eq:boxlinear}
	\bar{\mathcal{F}}_1(\Box)R=(R-6\lambda)\sum_{n=1}^\infty f_{1_{-n}}\bar{\lambda}^{-n}&=(R-6\lambda)\sum_{n=1}^\infty \tilde{f}_{1_{-n}}\biggl(\frac{M^2}{\bar{\lambda}}\biggr)^n=\bar{\mathcal{F}}_1(\bar{\lambda})(R-6\lambda)\,,
\end{align}
which we are going to discuss. An important remark is in order at this point. A  similar line of attack of the problem can  be adopted also at UV scale. See for example \cite{Spyridon, biswas2}.

\section{Dark energy as a nonlocal gravitational effect}
\label{sec:dark-energy}
Substituting the result \eqref{eq:boxlinear} into the nonlocal action \eqref{eq:action} yields
\begin{equation}
	\label{eq:actionansatz}
	S=\int d^{4}x\,\frac{\sqrt{-g}}{2}\bigl\{[M_\textup{P}^2-6\lambda\bar{\mathcal{F}}_1(\bar{\lambda})]R+\bar{\mathcal{F}}_1(\bar{\lambda})R^2\bigr\}.
\end{equation}
This expression shows that, for the particular solution \eqref{eq:scalefactor}, this \IDG~model represents a correction to the Starobinsky model, whose action can be expressed as
\begin{equation}
	\label{eq:starobinsky}
	S=\int d^{4}x\,\frac{\sqrt{-g}}{2}M_\textup{P}^2(R+\alpha R^2).
\end{equation}
The nonlocal effects, contained in the dimensionless coefficient $\bar{\mathcal{F}}_1(\bar{\lambda})$, act both as a correction to the gravitational constant (and therefore to Einstein’s \GR) and as a dynamical source for the $R^2$ term of Starobinsky gravity.

The coefficient $\bar{\mathcal{F}}_1(\bar{\lambda})$ remains to be determined. One possibility is to  compute all the coefficients $\tilde{f}_{1_{-n}}$ and then add them up as follows:
\begin{equation}
	\begin{split}
		\bar{\mathcal{F}}_1(\bar{\lambda})=\sum_{n=1}^\infty \tilde{f}_{1_{-n}}\biggl(\frac{M^2}{\bar{\lambda}}\biggr)^n&\approx-\frac{2M^2}{9\lambda}\tilde{f}_{1_{-1}}+\frac{4M^4}{81\lambda^2}\tilde{f}_{1_{-2}}-\frac{8M^6}{729\lambda^3}\tilde{f}_{1_{-3}}+\frac{16M^8}{6561\lambda^4}\tilde{f}_{1_{-4}}-\dots\\
		&\approx-0.0832\tilde{f}_{1_{-1}}+0.00693\tilde{f}_{1_{-2}}-0.000577\tilde{f}_{1_{-3}}+0.0000480\tilde{f}_{1_{-4}}-\dots.\,,
	\end{split}
\end{equation}
where we have retained three significant figures.
Substituting Eqs.\eqref{eq:scalefactor} and \eqref{eq:r1-r2} into  cosmological Eqs. \eqref{eq:cosmo1-matter} and \eqref{eq:cosmo2-matter} gives respectively
\begin{equation}
	\label{eq:constraint1}
	M_\textup{P}^2\lambda(2+3\tau)+48\bar{\mathcal{F}}_1(\bar{\lambda})\lambda^2(1+4\tau)+8\lambda^2\tau(8+9\tau)\sum_{n=1}^\infty nf_{1_{-n}}\bar{\lambda}^{-n}=0
\end{equation}
and
\begin{equation}
	\label{eq:constraint2}
	M_\textup{P}^2\lambda(1+2\tau)+24\bar{\mathcal{F}}_1(\bar{\lambda})\lambda^2(1+4\tau)+\frac{16}{3}\lambda^2\tau(4+9\tau)\sum_{n=1}^\infty nf_{1_{-n}}\bar{\lambda}^{-n}=\frac{1}{6}\Omega_m\rho_ce^{-\frac{3}{2}(\tau-\tau_0)}.
\end{equation}
The last two equations are constraints for the coefficients $\tilde{f}_{1_{-n}}$. First of all, it must be clarified why these equations contain the time variable $\tau$, whereas they should constrain some \emph{constant} quantities; this is the case because the ansatz \eqref{eq:ansatz} is only approximately satisfied, in particular it is satisfied only if a slowly-varying function of time (the generalised hypergeometric function ${}_2F_2$) is neglected: as a consequence, also the coefficients $\tilde{f}_{1_{-n}}$ should be some slowly-varying functions of time, in such a way that they can still be regarded as almost constant during the cosmological era of interest. Secondly, Eqs. \eqref{eq:constraint1} and \eqref{eq:constraint2} are two constraints whereby an infinite set of parameters, i.e. the coefficients $\tilde{f}_{1_{-n}}$, cannot be completely determined.

A theory with an infinite number of free parameters -- that cannot even be predicted -- is devoid of physical meaning. From a practical  point of view, it must exist the possibility to reformulate it in an alternative form containing a \emph{finite} number of parameters to avoid infinite fine-tunings.\footnote{An example for the occurrence of a similar rationale in Physics is given by physical systems described through the formalism of Statistical Mechanics.}

An approach is to calculate the dimensionless quantity $\bar{\mathcal{F}}_1(\bar{\lambda})$, since the nonlocal corrections can be rewritten in terms of this function only. It is a power series with unknown coefficients $\tilde{f}_{1_{-n}}$ and base $\frac{M^2}{\bar{\lambda}}$. If all of the infinite coefficients are of the same order of magnitude with alternating signs, it is a Leibniz series implying that $\bar{\mathcal{F}}_1(\bar{\lambda})$ is converging. We have to consider that  $\abs{\frac{M^2}{\bar{\lambda}}}=\frac{2M^2}{9\lambda}\sim\frac{1}{12}$. Secondly, cosmological Eqs. \eqref{eq:constraint1} and \eqref{eq:constraint2} constitute nonlinear constraints for $\bar{\mathcal{F}}_1(\bar{\lambda})$ and thus they cannot provide a general closed form for such coefficient.

A different approach to calculate $\bar{\mathcal{F}}_1(\bar{\lambda})$ is not considering coefficients $\tilde{f}_{1_{-n}}$ and going back to the nonlocal action \eqref{eq:actionansatz} obtained for the cosmological solution \eqref{eq:scalefactor}. Assuming the Starobinsky gravity  \eqref{eq:starobinsky}, one can define
\begin{equation}
	\bar{\mathcal{F}}_1(\bar{\lambda})\equiv M_\textup{P}^2\alpha,
\end{equation}
so that the action \eqref{eq:actionansatz} can be put into a more concise form:
\begin{equation}
	S=\int d^{4}x\,\frac{\sqrt{-g}}{2}M_\textup{P}^2[(1-6\alpha\lambda)R+\alpha R^2].
\end{equation}
The field equations obtained from the variational principle are a straightforward generalisation of those of Starobinsky gravity, that is:
\begin{equation}
	\label{eq:eq-motion-starobinsky}
	(1-6\alpha\lambda)G_{\mu\nu}+2\alpha\biggl(RR_{\mu\nu}-\frac{1}{4}g_{\mu\nu}R^2+g_{\mu\nu}\Box R-\nabla_\mu\nabla_\nu R\biggr)=M_\textup{P}^{-2}T_{\mu\nu}.
\end{equation}
Also in this case, we are assuming that the coefficients $\tilde{f}_{1_{-n}}$ are slowly-varying functions of time, which means that the quantity $\alpha$ can be considered as effectively constant during the cosmological era of interest.

The trace equation of Eqs. \eqref{eq:eq-motion-starobinsky} is
\begin{equation}
	6\alpha\Box R-(1-6\alpha\lambda)R=M_\textup{P}^{-2}T.
\end{equation}
Substituting expressions \eqref{eq:Ransatz} and \eqref{eq:boxRansatz}, one  finds
\begin{equation}
	\alpha=\frac{M_\textup{P}^{-2}T+R}{6(\lambda+\Box)R}=\frac{\frac{\Omega_m\rho_c}{6M_\textup{P}^2\lambda}e^{-\frac{3}{2}(\tau-\tau_0)}-2\tau-1}{6\lambda(10\tau+3)},
\end{equation}
which, when evaluated at the present cosmic time, yields the following result:
\begin{equation}
	\alpha=\frac{\frac{\Omega_m\rho_c}{6M_\textup{P}^2\lambda}-2\tau_0-1}{6\lambda(10\tau_0+3)},
\end{equation}
that is $\alpha=-1.37\cdot10^{82}$\,GeV\textsuperscript{-2}.\footnote{As a check that $\alpha$ can indeed be considered as almost constant in the cosmological era of interest, we note that the relative error between its value at $\tau_0$ and its value at $\tau_*$ is less than 3\%.} This implies that
\begin{equation}
	\bar{\mathcal{F}}_1(\bar{\lambda})=\frac{\frac{\Omega_m\rho_c}{6\lambda}-M_\textup{P}^2(2\tau_0+1)}{6\lambda(10\tau_0+3)}
\end{equation}
and so $\bar{\mathcal{F}}_1(\bar{\lambda})=-8.14\cdot10^{118}$.

Let us discuss this result in view of the nonlocal action \eqref{eq:actionansatz}. The coefficient of the first term, the one proportional to $R$, is $M_\textup{P}^2-6\lambda\bar{\mathcal{F}}_1(\bar{\lambda})=7.26\cdot10^{36}$\,GeV\textsuperscript{2}$=1.22\,M_\textup{P}^2$; in particular, the result $\bar{\mathcal{F}}_1(\bar{\lambda})<0$ implies a decrease of the effective gravitational constant in the current cosmological era, i.e. $G\rightarrow G_\textup{eff}=G/1.22=82\%\,G$, thereby possibly explaining dark energy away as the manifestation of nonlocal gravitational effects on cosmological scales. These nonlocal effects can thus be regarded either as repulsive gravity (in the sense of spacetime curvature) or as negative pressure (in the sense of matter sources).

The coefficient of the second term, the one proportional to $R^2$, is $\bar{\mathcal{F}}_1(\bar{\lambda})$, whose negative sign could hint at the presence of ghosts in the theory: that is the case in a Minkowski background \cite{biswas:propagator}, but it has yet to be demonstrated in a curved spacetime as well. Furthermore, in the Starobinsky model, the parameter $\alpha$ corresponds to a mass being $m=1/\sqrt{-6\alpha}$, yielding $m\sim10^{-42}$\,GeV$\sim H_0$, so the sum of the nonlocal series can be physically interpreted as proportional to the mass of a (very light) scalar field.

\section{Discussion and Conclusions}
\label{sec:conclusions}
The present analysis shows that it is possible to describe  dark energy  in the context of \IDG, considered as an effective theory of gravity with nonlocal terms. This represents an alternative picture to the addition of  cosmological constant  in \GR~or to the introduction of some unknown scalar fields. In particular, the considered phenomenological model provides an accelerated expansion of the late Universe without affecting the evolution of the previous cosmological eras, as the nonlocal effects are significant only in the IR limit.

Interestingly, the model also predicts a reduction of the effective gravitational constant in the current cosmological era: therefore, it could be possible that, after the matter-dominated era, the evolution of the Universe is driven by the emergence of nonlocal gravitational effects, which remain `hidden' until the Universe has cooled enough. See also \cite{vanPutten2}. This conclusion points out the fact that \GR~is valid with great accuracy in the energy interval between the IR limit, set by $M$, where nonlocal effects become dominant, and the UV limit, set by $M_\textup{P}$, where quantum effects take over. At cosmological scales, results from the \textit{JWST} and  \textit{Euclid} missions could be indicative in this direction  \cite{JWST,vanPutten3}. The very recent DESI results could be extremely interesting in this perspective \cite{DESI}.

Another important issue is related to the fact that,  in our approach,  it is natural to recover  the Starobinsky model, improved with  non-local corrections. This scenario can produce  interesting inflationary regimes. See Refs. \cite{Koshelev4,Koshelev2}. At inflationary epochs, nonlocal effects produce a  blue tilted tensor spectrum with small non-Gaussianities   \cite{Koshelev1,Koshelev3}.  In this perspective, the present late time IR analysis   could be  useful to realize a straightforward  unification between  inflation and  dark energy era. This topic will be developed in a forthcoming paper.

Finally,  nonlocal terms can be probed  by Lunar Laser Ranging  results pointing out  time-variations of  the gravitational coupling. This approach seems to  rule out some nonlocal gravity models as discussed in Ref. \cite{Belgacem}. In particular, the Deser-Woodard models would be not viable   while models considering the extraction of the transverse part from the tensor $(\Box^{-1}G_{\mu\nu})$ result dynamically consistent. In the present discussion, we have taken into cosmological scales where nonlocal terms drive the IR behavior. In future investigations, these models will be tested at Solar System scales with the aim to achieve a general picture of nonlocal dynamics.

Moreover, one should not overlook that the present analysis indicates that  \IDG~can be predictive without resorting to a truncation of its infinitely many derivative terms, which play a crucial role for the quantization of the theory \cite{biswas:towards}. Besides the applications in the IR limit, the formalism outlined here, as pointed out above,  can be applied also in the UV limit, considering, in particular, the Starobinsky inflation and the primordial cosmological perturbations.

A general remark is important before concluding the discussion.  The effect of having eras in the evolution of the  Universe is  intriguing and, very likely, is the key aspect to achieve a whole cosmic history. However,  mechanisms similar to those discussed here can be considered taking into account other  extended gravities like $f(R)$ or Gauss-Bonnet gravity.    The general idea is that geometric corrections to General Relativity can trigger the transitions to the various epochs. Some interesting examples  are reported in Refs. \cite{VasA,VasB,VasC}.

\section*{Aknowledgements}
This paper is based upon work from  the COST Action CA21136, ``Addressing observational tensions in cosmology with systematics and fundamental physics'' (CosmoVerse) supported by COST (European Cooperation in Science and Technology). SC and GM acknowledge the support of the  Istituto Nazionale di Fisica Nucleare, Sez. di Napoli, Iniziativa Specifica QGSKY.

\appendix*
\section{Derivation of the nonlocal cosmological equations}
\label{sec:appendix}
\setcounter{equation}{0}
\renewcommand{\theequation}{A\arabic{equation}}
Here we provide an in-depth calculation of nonlocal cosmological Eqs. \eqref{eq:cosmo1} and \eqref{eq:cosmo2}. The nonzero components of the Ricci tensor in the flat \FLRW~metric \eqref{eq:metric} are
\begin{equation}
	R_{00}=-3\frac{\ddot{a}}{a},\qquad R_{11}=R_{22}=R_{33}=a\ddot{a}+2\dot{a}^2.
\end{equation}
Thus, making use of the expression \eqref{eq:ricciscalar} for the Ricci scalar, we have that the `00'-component and the `11'-component of the Einstein tensor are given respectively by
\begin{equation}
	G_{00}=R_{00}-\frac{1}{2}g_{00}R=3\frac{\dot{a}^2}{a^2},\qquad G_{11}=R_{11}-\frac{1}{2}g_{11}R=-2a\ddot{a}-\dot{a}^2.
\end{equation}
The stress-energy tensor for a perfect fluid in the metric \eqref{eq:metric} is given by
\begin{equation}
	T_\mu^\nu=\diag\Bigl(-\rho(t),p(t),p(t),p(t)\Bigr),
\end{equation}
so its relevant components are
\begin{equation}
	T_{00}=\rho,\qquad T_{11}=a^2p.
\end{equation}
Furthermore, considering that
\begin{align}
	\bar{\mathcal{F}}_1(\Box)R&=6\sum_{n=1}^\infty f_{1_{-n}}\Box^{-n}\biggl(\frac{a\ddot{a}+\dot{a}^2}{a^2}\biggr),\\
	\label{eq:theta1}
	\Theta_{\mu\nu}^1&=36\delta_\mu^0\delta_\nu^0\sum_{n=1}^\infty f_{1_{-n}}\sum_{l=0}^{n-1}\partial_0\Box^{-l-1}\biggl(\frac{a\ddot{a}+\dot{a}^2}{a^2}\biggr)\partial_0\Box^{-n+l}\biggl(\frac{a\ddot{a}+\dot{a}^2}{a^2}\biggr),\\
	\Theta_\sigma^{1\sigma}&=g^{\mu\nu}\Theta_{\mu\nu}^1=-36\sum_{n=1}^\infty f_{1_{-n}}\sum_{l=0}^{n-1}\partial_0\Box^{-l-1}\biggl(\frac{a\ddot{a}+\dot{a}^2}{a^2}\biggr)\partial_0\Box^{-n+l}\biggl(\frac{a\ddot{a}+\dot{a}^2}{a^2}\biggr),\\
	\bar{\Theta}^1&=36\sum_{n=1}^\infty f_{1_{-n}}\sum_{l=0}^{n-1}\Box^{-l-1}\biggl(\frac{a\ddot{a}+\dot{a}^2}{a^2}\biggr)\Box^{-n+l+1}\biggl(\frac{a\ddot{a}+\dot{a}^2}{a^2}\biggr),\\
	\partial_0R&=6\frac{a^2\dddot{a}+a\dot{a}\ddot{a}-2\dot{a}^3}{a^3},\\
	\label{eq:theta2}
	\Box R&=-6a^{-3}\partial_0\biggl(a^3\partial_0\frac{a\ddot{a}+\dot{a}^2}{a^2}\biggr)=-6\frac{a\ddot{a}^2-5\dot{a}^2\ddot{a}+3a\dot{a}\dddot{a}+a^2\ddddot{a}}{a^3},
\end{align}
the `00'-component and the `11'-component of field Eqs. \eqref{eq:eq-motion} can be rewritten respectively as
\begin{subequations}
	\begin{align}
		T_{00}&=M_\textup{P}^2G_{00}+2G_{00}\bar{\mathcal{F}}_1(\Box)R+\frac{1}{2}g_{00}R\bar{\mathcal{F}}_1(\Box)R+\Theta_{00}^1-\frac{1}{2}g_{00}\bigl(\Theta_\sigma^{1\sigma}+\bar{\Theta}^1\bigr),\\
		T_{11}&=M_\textup{P}^2G_{11}+2G_{11}\bar{\mathcal{F}}_1(\Box)R+\frac{1}{2}g_{11}R\bar{\mathcal{F}}_1(\Box)R+2g_{11}\Box\bar{\mathcal{F}}_1(\Box)R-\frac{1}{2}g_{11}\bigl(\Theta_\sigma^{1\sigma}+\bar{\Theta}^1\bigr)
	\end{align}
\end{subequations}
or, more explicitly, as
\begin{equation}
	\label{eq:cosmo1temp}
	\begin{split}
		\rho&=3M_\textup{P}^2\frac{\dot{a}^2}{a^2}+18\biggl(-\frac{\ddot{a}}{a}+\frac{\dot{a}^2}{a^2}\biggr)\sum_{n=1}^\infty f_{1_{-n}}\Box^{-n}\biggl(\frac{a\ddot{a}+\dot{a}^2}{a^2}\biggr)\\
		&+18\sum_{n=1}^\infty f_{1_{-n}}\sum_{l=0}^{n-1}\partial_0\Box^{-l-1}\biggl(\frac{a\ddot{a}+\dot{a}^2}{a^2}\biggr)\partial_0\Box^{-n+l}\biggl(\frac{a\ddot{a}+\dot{a}^2}{a^2}\biggr)+18\sum_{n=1}^\infty f_{1_{-n}}\sum_{l=0}^{n-1}\Box^{-l-1}\biggl(\frac{a\ddot{a}+\dot{a}^2}{a^2}\biggr)\Box^{-n+l+1}\biggl(\frac{a\ddot{a}+\dot{a}^2}{a^2}\biggr)
	\end{split}
\end{equation}
and
\begin{equation}
	\label{eq:cosmo2temp}
	\begin{split}
		p&=-M_\textup{P}^2\biggl(2\frac{\ddot{a}}{a}+\frac{\dot{a}^2}{a^2}\biggr)+6\biggl[-\frac{\ddot{a}}{a}+\frac{\dot{a}^2}{a^2}-2a^{-3}\partial_0(a^3\partial_0)\biggr]\sum_{n=1}^\infty f_{1_{-n}}\Box^{-n}\biggl(\frac{a\ddot{a}+\dot{a}^2}{a^2}\biggr)\\
		&+18\sum_{n=1}^\infty f_{1_{-n}}\sum_{l=0}^{n-1}\partial_0\Box^{-l-1}\biggl(\frac{a\ddot{a}+\dot{a}^2}{a^2}\biggr)\partial_0\Box^{-n+l}\biggl(\frac{a\ddot{a}+\dot{a}^2}{a^2}\biggr)-18\sum_{n=1}^\infty f_{1_{-n}}\sum_{l=0}^{n-1}\Box^{-l-1}\biggl(\frac{a\ddot{a}+\dot{a}^2}{a^2}\biggr)\Box^{-n+l+1}\biggl(\frac{a\ddot{a}+\dot{a}^2}{a^2}\biggr).
	\end{split}
\end{equation}
Using the equation of state, $p=w\rho$, Eqs. \eqref{eq:cosmo1temp} and \eqref{eq:cosmo2temp} can be combined into a single integro-differential equation for the scale factor:
\begin{equation}
	\label{eq:cosmo3temp}
	\begin{split}
		&M_\textup{P}^2\biggl[2\frac{\ddot{a}}{a}+(3w+1)\frac{\dot{a}^2}{a^2}\biggr]-6\biggl[(3w-1)\biggl(\frac{\ddot{a}}{a}-\frac{\dot{a}^2}{a^2}\biggr)-2a^{-3}\partial_0(a^3\partial_0)\biggr]\sum_{n=1}^\infty f_{1_{-n}}\Box^{-n}\biggl(\frac{a\ddot{a}+\dot{a}^2}{a^2}\biggr)\\
		&+18(w-1)\sum_{n=1}^\infty f_{1_{-n}}\sum_{l=0}^{n-1}\partial_0\Box^{-l-1}\biggl(\frac{a\ddot{a}+\dot{a}^2}{a^2}\biggr)\partial_0\Box^{-n+l}\biggl(\frac{a\ddot{a}+\dot{a}^2}{a^2}\biggr)\\
		&+18(w+1)\sum_{n=1}^\infty f_{1_{-n}}\sum_{l=0}^{n-1}\Box^{-l-1}\biggl(\frac{a\ddot{a}+\dot{a}^2}{a^2}\biggr)\Box^{-n+l+1}\biggl(\frac{a\ddot{a}+\dot{a}^2}{a^2}\biggr)=0.
	\end{split}
\end{equation}
Carrying out the same substitutions for trace Eq. \eqref{eq:trace-eq}, one obtains
\begin{equation}
	\label{eq:cosmo4temp}
	\begin{split}
		M_\textup{P}^2\frac{a\ddot{a}+\dot{a}^2}{a^2}&+6a^{-3}\partial_0(a^3\partial_0)\sum_{n=1}^\infty f_{1_{-n}}\Box^{-n}\biggl(\frac{a\ddot{a}+\dot{a}^2}{a^2}\biggr)-6\sum_{n=1}^\infty f_{1_{-n}}\sum_{l=0}^{n-1}\partial_0\Box^{-l-1}\biggl(\frac{a\ddot{a}+\dot{a}^2}{a^2}\biggr)\partial_0\Box^{-n+l}\biggl(\frac{a\ddot{a}+\dot{a}^2}{a^2}\biggr)\\
		&+12\sum_{n=1}^\infty f_{1_{-n}}\sum_{l=0}^{n-1}\Box^{-l-1}\biggl(\frac{a\ddot{a}+\dot{a}^2}{a^2}\biggr)\Box^{-n+l+1}\biggl(\frac{a\ddot{a}+\dot{a}^2}{a^2}\biggr)=\frac{1-3w}{6}\rho.
	\end{split}
\end{equation}
The nonlocal operators appearing in Eqs. \eqref{eq:cosmo3temp} and \eqref{eq:cosmo4temp} can be evaluated as follows:
\begin{align}
	\label{eq:box-1}
	\biggl[\Box^{-1}\biggl(\frac{a\ddot{a}+\dot{a}^2}{a^2}\biggr)\biggr](t)&=-\int_0^t dt'\,a^{-3}(t')\int_0^{t'} dt''\,(a^2\ddot{a}+a\dot{a}^2)(t''),\\
	\label{eq:box-1-partial0}
	\biggl[\partial_0\Box^{-1}\biggl(\frac{a\ddot{a}+\dot{a}^2}{a^2}\biggr)\biggr](t)&=\biggl[\Box^{-1}\biggl(\frac{a^2\dddot{a}+a\dot{a}\ddot{a}-2\dot{a}^3}{a^3}\biggr)\biggr](t)=-\int_0^t dt'\,a^{-3}(t')\int_0^{t'} dt''\,(a^2\dddot{a}+a\dot{a}\ddot{a}-2\dot{a}^3)(t'').
\end{align}
By recursive application of the operator $\Box^{-1}$, from Eqs. \eqref{eq:box-1} and \eqref{eq:box-1-partial0}, one finds respectively that
\begin{align}
	\label{eq:box-n}
	&\biggl[\Box^{-n}\biggl(\frac{a\ddot{a}+\dot{a}^2}{a^2}\biggr)\biggr](t)\nonumber\\
	&=(-1)^n\int_0^t dt'\,a^{-3}(t')\int_0^{t'} dt''\,a^3(t'')\dots\int_0^{t^{(2n-2)}} dt^{(2n-1)}\,a^{-3}\bigl(t^{(2n-1)}\bigr)\int_0^{t^{(2n-1)}} dt^{(2n)}\,(a^2\ddot{a}+a\dot{a}^2)\bigl(t^{(2n)}\bigr),\\
	\label{eq:box-n-partial0}
	&\biggl[\partial_0\Box^{-n}\biggl(\frac{a\ddot{a}+\dot{a}^2}{a^2}\biggr)\biggr](t)=\biggl[\Box^{-n}\biggl(\frac{a^2\dddot{a}+a\dot{a}\ddot{a}-2\dot{a}^3}{a^3}\biggr)\biggr](t)\nonumber\\
	&=(-1)^n\int_0^t dt'\,a^{-3}(t')\int_0^{t'} dt''\,a^3(t'')\dots\int_0^{t^{(2n-2)}} dt^{(2n-1)}\,a^{-3}\bigl(t^{(2n-1)}\bigr)\int_0^{t^{(2n-1)}} dt^{(2n)}\,(a^2\dddot{a}+a\dot{a}\ddot{a}-2\dot{a}^3)\bigl(t^{(2n)}\bigr),
\end{align}
where both expressions contain $2n$ time integrals with $2n$ variables of integration $t',t'',\dots,t^{(2n-1)},t^{(2n)}$. Finally, substituting the results \eqref{eq:box-n} and \eqref{eq:box-n-partial0} into Eqs. \eqref{eq:cosmo3temp} and \eqref{eq:cosmo4temp} yields
\begin{equation}
	\label{eq:cosmo1}
	\begin{split}
		&0=M_\textup{P}^2\biggl[2\frac{\ddot{a}}{a}+(3w+1)\frac{\dot{a}^2}{a^2}\biggr]-6\biggl[(3w-1)\biggl(\frac{\ddot{a}}{a}-\frac{\dot{a}^2}{a^2}\biggr)-2a^{-3}\partial_0(a^3\partial_0)\biggr]\Biggl\{\sum_{n=1}^\infty (-1)^nf_{1_{-n}}\\
		&\cdot\Biggl[\int_0^t dt'\,a^{-3}(t')\int_0^{t'} dt''\,a^3(t'')\dots\int_0^{t^{(2n-2)}} dt^{(2n-1)}\,a^{-3}\bigl(t^{(2n-1)}\bigr)\int_0^{t^{(2n-1)}} dt^{(2n)}\,(a^2\ddot{a}+a\dot{a}^2)\bigl(t^{(2n)}\bigr)\Biggr]\Biggr\}\\
		&+18(w-1)\Biggl\{\sum_{n=1}^\infty(-1)^{n+1}f_{1_{-n}}\\
		&\cdot\sum_{l=0}^{n-1}\Biggl[\int_0^t dt'\,a^{-3}(t')\int_0^{t'} dt''\,a^3(t'')\dots\int_0^{t^{(2l)}} dt^{(2l+1)}\,a^{-3}\bigl(t^{(2l+1)}\bigr)\int_0^{t^{(2l+1)}} dt^{(2l+2)}\,(a^2\dddot{a}+a\dot{a}\ddot{a}-2\dot{a}^3)\bigl(t^{(2l+2)}\bigr)\Biggr]\\
		&\cdot\Biggl[\int_0^t dt'\,a^{-3}(t')\int_0^{t'} dt''\,a^3(t'')\dots\int_0^{t^{(2n-2l-2)}} dt^{(2n-2l-1)}\,a^{-3}\bigl(t^{(2n-2l-1)}\bigr)\int_0^{t^{(2n-2l-1)}} dt^{(2n-2l)}\,(a^2\dddot{a}+a\dot{a}\ddot{a}-2\dot{a}^3)\bigl(t^{(2n-2l)}\bigr)\Biggr]\Biggr\}\\
		&+18(w+1)\Biggl\{\sum_{n=1}^\infty(-1)^{n+1}f_{1_{-n}}\\
		&\cdot\sum_{l=0}^{n-1}\Biggl[\int_0^t dt'\,a^{-3}(t')\int_0^{t'} dt''\,a^3(t'')\dots\int_0^{t^{(2l)}} dt^{(2l+1)}\,a^{-3}\bigl(t^{(2l+1)}\bigr)\int_0^{t^{(2l+1)}} dt^{(2l+2)}\,(a^2\ddot{a}+a\dot{a}^2)\bigl(t^{(2l+2)}\bigr)\Biggr]\\
		&\cdot\Biggl[\int_0^t dt'\,a^{-3}(t')\int_0^{t'} dt''\,a^3(t'')\dots\int_0^{t^{(2n-2l-4)}} dt^{(2n-2l-3)}\,a^{-3}\bigl(t^{(2n-2l-3)}\bigr)\int_0^{t^{(2n-2l-3)}} dt^{(2n-2l-2)}\,(a^2\ddot{a}+a\dot{a}^2)\bigl(t^{(2n-2l-2)}\bigr)\Biggr]\Biggr\}
	\end{split}
\end{equation}
and
\begin{equation}
	\label{eq:cosmo2}
	\begin{split}
		&\frac{1-3w}{6}\rho=M_\textup{P}^2\frac{a\ddot{a}+\dot{a}^2}{a^2}+6a^{-3}\partial_0(a^3\partial_0)\Biggl\{\sum_{n=1}^\infty (-1)^nf_{1_{-n}}\\
		&\cdot\Biggl[\int_0^t dt'\,a^{-3}(t')\int_0^{t'} dt''\,a^3(t'')\dots\int_0^{t^{(2n-2)}} dt^{(2n-1)}\,a^{-3}\bigl(t^{(2n-1)}\bigr)\int_0^{t^{(2n-1)}} dt^{(2n)}\,(a^2\ddot{a}+a\dot{a}^2)\bigl(t^{(2n)}\bigr)\Biggr]\Biggr\}\\
		&-6\Biggl\{\sum_{n=1}^\infty(-1)^{n+1}f_{1_{-n}}\\
		&\cdot\sum_{l=0}^{n-1}\Biggl[\int_0^t dt'\,a^{-3}(t')\int_0^{t'} dt''\,a^3(t'')\dots\int_0^{t^{(2l)}} dt^{(2l+1)}\,a^{-3}\bigl(t^{(2l+1)}\bigr)\int_0^{t^{(2l+1)}} dt^{(2l+2)}\,(a^2\dddot{a}+a\dot{a}\ddot{a}-2\dot{a}^3)\bigl(t^{(2l+2)}\bigr)\Biggr]\\
		&\cdot\Biggl[\int_0^t dt'\,a^{-3}(t')\int_0^{t'} dt''\,a^3(t'')\dots\int_0^{t^{(2n-2l-2)}} dt^{(2n-2l-1)}\,a^{-3}\bigl(t^{(2n-2l-1)}\bigr)\int_0^{t^{(2n-2l-1)}} dt^{(2n-2l)}\,(a^2\dddot{a}+a\dot{a}\ddot{a}-2\dot{a}^3)\bigl(t^{(2n-2l)}\bigr)\Biggr]\Biggr\}\\
		&+12\Biggl\{\sum_{n=1}^\infty(-1)^{n+1}f_{1_{-n}}\\
		&\cdot\sum_{l=0}^{n-1}\Biggl[\int_0^t dt'\,a^{-3}(t')\int_0^{t'} dt''\,a^3(t'')\dots\int_0^{t^{(2l)}} dt^{(2l+1)}\,a^{-3}\bigl(t^{(2l+1)}\bigr)\int_0^{t^{(2l+1)}} dt^{(2l+2)}\,(a^2\ddot{a}+a\dot{a}^2)\bigl(t^{(2l+2)}\bigr)\Biggr]\\
		&\cdot\Biggl[\int_0^t dt'\,a^{-3}(t')\int_0^{t'} dt''\,a^3(t'')\dots\int_0^{t^{(2n-2l-4)}} dt^{(2n-2l-3)}\,a^{-3}\bigl(t^{(2n-2l-3)}\bigr)\int_0^{t^{(2n-2l-3)}} dt^{(2n-2l-2)}\,(a^2\ddot{a}+a\dot{a}^2)\bigl(t^{(2n-2l-2)}\bigr)\Biggr]\Biggr\}.
	\end{split}
\end{equation}
In the matter-dominated era, nonlocal cosmological Eqs. \eqref{eq:cosmo1} and \eqref{eq:cosmo2} become respectively
\begin{equation}
	\begin{split}
		&0=M_\textup{P}^2\biggl[2\frac{\ddot{a}}{a}+(3w+1)\frac{\dot{a}^2}{a^2}\biggr]-6\biggl[-\frac{\ddot{a}}{a}+\frac{\dot{a}^2}{a^2}-2a^{-3}\partial_0(a^3\partial_0)\biggr]\Biggl\{\sum_{n=1}^\infty (-1)^nf_{1_{-n}}\\
		&\cdot\Biggl[\int_0^t dt'\,a^{-3}(t')\int_0^{t'} dt''\,a^3(t'')\dots\int_0^{t^{(2n-2)}} dt^{(2n-1)}\,a^{-3}\bigl(t^{(2n-1)}\bigr)\int_0^{t^{(2n-1)}} dt^{(2n)}\,(a^2\ddot{a}+a\dot{a}^2)\bigl(t^{(2n)}\bigr)\Biggr]\Biggr\}\\
		&-18\Biggl\{\sum_{n=1}^\infty(-1)^{n+1}f_{1_{-n}}\\
		&\cdot\sum_{l=0}^{n-1}\Biggl[\int_0^t dt'\,a^{-3}(t')\int_0^{t'} dt''\,a^3(t'')\dots\int_0^{t^{(2l)}} dt^{(2l+1)}\,a^{-3}\bigl(t^{(2l+1)}\bigr)\int_0^{t^{(2l+1)}} dt^{(2l+2)}\,(a^2\dddot{a}+a\dot{a}\ddot{a}-2\dot{a}^3)\bigl(t^{(2l+2)}\bigr)\Biggr]\\
		&\cdot\Biggl[\int_0^t dt'\,a^{-3}(t')\int_0^{t'} dt''\,a^3(t'')\dots\int_0^{t^{(2n-2l-2)}} dt^{(2n-2l-1)}\,a^{-3}\bigl(t^{(2n-2l-1)}\bigr)\int_0^{t^{(2n-2l-1)}} dt^{(2n-2l)}\,(a^2\dddot{a}+a\dot{a}\ddot{a}-2\dot{a}^3)\bigl(t^{(2n-2l)}\bigr)\Biggr]\Biggr\}\\
		&+18\Biggl\{\sum_{n=1}^\infty(-1)^{n+1}f_{1_{-n}}\\
		&\cdot\sum_{l=0}^{n-1}\Biggl[\int_0^t dt'\,a^{-3}(t')\int_0^{t'} dt''\,a^3(t'')\dots\int_0^{t^{(2l)}} dt^{(2l+1)}\,a^{-3}\bigl(t^{(2l+1)}\bigr)\int_0^{t^{(2l+1)}} dt^{(2l+2)}\,(a^2\ddot{a}+a\dot{a}^2)\bigl(t^{(2l+2)}\bigr)\Biggr]\\
		&\cdot\Biggl[\int_0^t dt'\,a^{-3}(t')\int_0^{t'} dt''\,a^3(t'')\dots\int_0^{t^{(2n-2l-4)}} dt^{(2n-2l-3)}\,a^{-3}\bigl(t^{(2n-2l-3)}\bigr)\int_0^{t^{(2n-2l-3)}} dt^{(2n-2l-2)}\,(a^2\ddot{a}+a\dot{a}^2)\bigl(t^{(2n-2l-2)}\bigr)\Biggr]\Biggr\}
	\end{split}
\end{equation}
and
\begin{equation}
	\begin{split}
		&\frac{1}{6}\Omega_m\rho_ca_0^3a^{-3}=M_\textup{P}^2\frac{a\ddot{a}+\dot{a}^2}{a^2}+6a^{-3}\partial_0(a^3\partial_0)\Biggl\{\sum_{n=1}^\infty (-1)^nf_{1_{-n}}\\
		&\cdot\Biggl[\int_0^t dt'\,a^{-3}(t')\int_0^{t'} dt''\,a^3(t'')\dots\int_0^{t^{(2n-2)}} dt^{(2n-1)}\,a^{-3}\bigl(t^{(2n-1)}\bigr)\int_0^{t^{(2n-1)}} dt^{(2n)}\,(a^2\ddot{a}+a\dot{a}^2)\bigl(t^{(2n)}\bigr)\Biggr]\Biggr\}\\
		&-6\Biggl\{\sum_{n=1}^\infty(-1)^{n+1}f_{1_{-n}}\\
		&\cdot\sum_{l=0}^{n-1}\Biggl[\int_0^t dt'\,a^{-3}(t')\int_0^{t'} dt''\,a^3(t'')\dots\int_0^{t^{(2l)}} dt^{(2l+1)}\,a^{-3}\bigl(t^{(2l+1)}\bigr)\int_0^{t^{(2l+1)}} dt^{(2l+2)}\,(a^2\dddot{a}+a\dot{a}\ddot{a}-2\dot{a}^3)\bigl(t^{(2l+2)}\bigr)\Biggr]\\
		&\cdot\Biggl[\int_0^t dt'\,a^{-3}(t')\int_0^{t'} dt''\,a^3(t'')\dots\int_0^{t^{(2n-2l-2)}} dt^{(2n-2l-1)}\,a^{-3}\bigl(t^{(2n-2l-1)}\bigr)\int_0^{t^{(2n-2l-1)}} dt^{(2n-2l)}\,(a^2\dddot{a}+a\dot{a}\ddot{a}-2\dot{a}^3)\bigl(t^{(2n-2l)}\bigr)\Biggr]\Biggr\}\\
		&+12\Biggl\{\sum_{n=1}^\infty(-1)^{n+1}f_{1_{-n}}\\
		&\cdot\sum_{l=0}^{n-1}\Biggl[\int_0^t dt'\,a^{-3}(t')\int_0^{t'} dt''\,a^3(t'')\dots\int_0^{t^{(2l)}} dt^{(2l+1)}\,a^{-3}\bigl(t^{(2l+1)}\bigr)\int_0^{t^{(2l+1)}} dt^{(2l+2)}\,(a^2\ddot{a}+a\dot{a}^2)\bigl(t^{(2l+2)}\bigr)\Biggr]\\
		&\cdot\Biggl[\int_0^t dt'\,a^{-3}(t')\int_0^{t'} dt''\,a^3(t'')\dots\int_0^{t^{(2n-2l-4)}} dt^{(2n-2l-3)}\,a^{-3}\bigl(t^{(2n-2l-3)}\bigr)\int_0^{t^{(2n-2l-3)}} dt^{(2n-2l-2)}\,(a^2\ddot{a}+a\dot{a}^2)\bigl(t^{(2n-2l-2)}\bigr)\Biggr]\Biggr\}.
	\end{split}
\end{equation}


\begin{thebibliography}{99}

\bibitem{Abdalla}
E.~Abdalla, G.~Franco Abell\'an, A.~Aboubrahim, A.~Agnello, O.~Akarsu, Y.~Akrami, G.~Alestas, D.~Aloni, L.~Amendola and L.~A.~Anchordoqui, \textit{et al.},
Cosmology intertwined: A review of the particle physics, astrophysics, and cosmology associated with the cosmological tensions and anomalies,
JHEAp \textbf{34}, 49-211 (2022).

\bibitem{goroff:ultraviolet}
M. H. Goroff, A. Sagnotti, The ultraviolet behavior of Einstein gravity, Nucl. Phys. B \textbf{266}, 709--736 (1986).

\bibitem{Starobinsky}
A.~A.~Starobinsky,
A New Type of Isotropic Cosmological Models Without Singularity,
Phys. Lett. B \textbf{91}, 99-102 (1980).

\bibitem{linde:inflationary}
A. D. Linde, Inflationary Cosmology, \textit{Inflationary Cosmology}, 1--54 (2007).

\bibitem{guth:inflationary}
A. H. Guth, Inflationary universe: A possible solution to the horizon and flatness problems, Phys. Rev. D \textbf{23}, 347 (1981).

\bibitem{linde:new}
A. D. Linde, A new inflationary universe scenario: a possible solution of the horizon, flatness, homogeneity, isotropy and primordial monopole problems, Phys. Lett. B \textbf{108}, 389--393 (1982).

\bibitem{linde:chaotic}
A. D. Linde, Chaotic inflation, Phys. Lett. B \textbf{129}, 177--181 (1983).

\bibitem{hartle:wave}
J. B. Hartle, S. W. Hawking, Wave function of the Universe, Phys. Rev. D \textbf{28} (1983).

\bibitem{vilenkin:quantum}
A. Vilenkin, Quantum creation of universes, Phys. Rev. D \textbf{30}, 509 (1984).

\bibitem{halliwell:introductory}
J. J. Halliwell, Introductory lectures on Quantum Cosmology, \textit{Quantum Cosmology and Baby Universes}, 159--243 (1991).

\bibitem{vanPutten1}
M.~H.~P.~M.~van Putten,
$H_0$-tension in the classical limit of Big Bang quantum cosmology,\\
{\it  32$^{nd}$ Texas Symp. Rel. Astroph., Shanghai, Session 7 (2023)}
[arXiv:2403.10865 [astro-ph.CO]] (2024).


\bibitem{planck:results}
N. Aghanim \textit{et al.}, Planck Collaboration, Planck 2018 results. VI. Cosmological parameters, Astron. Astrophys. \textbf{641}, A6 (2020).

\bibitem{weinberg:cosmological}
S. Weinberg, The Cosmological Constant Problem, Rev. Mod. Phys. \textbf{61}, 1--23 (1989).


\bibitem{conroy:generalised}
A. Conroy, T. Koivisto, A. Mazumdar, A. Teimouri, Generalised Quadratic Curvature, Non-Local Infrared Modifications of Gravity and Newtonian Potentials, Class. Quant. Grav. \textbf{32}, 015024 (2014).

\bibitem{vanPutten2}
M.~H.~P.~M.~van Putten,
Entropic constraint on cosmic variation of Planck mass and the Boltzmann constant,
Res. in Phys. \textbf{57}, 107425 (2024).

\bibitem{extended}
S.~Capozziello and M.~Francaviglia,
Extended Theories of Gravity and their Cosmological and Astrophysical Applications,
Gen. Rel. Grav. \textbf{40}, 357-420 (2008).


\bibitem{oikonomou}
S.~Nojiri, S.~D.~Odintsov and V.~K.~Oikonomou,
Modified Gravity Theories on a Nutshell: Inflation, Bounce and Late-time Evolution,
Phys. Rept. \textbf{692}, 1-104 (2017).

\bibitem{Clifton}  
T. Clifton, P. G. Ferreira, A. Padilla, C.  Skordis,
Modified Gravity and Cosmology, 
Physics Reports \textbf{513},  1-189 (2012).


\bibitem{SD}
S.  Nojiri, S. D. Odintsov, 
Unified cosmic history in modified gravity: from F(R) theory to Lorentz non-invariant models,
Phys.Rept.{\bf 505}, 59-144 (2011).

\bibitem{Rept}
S. Capozziello, M. De Laurentis,
Extended Theories of Gravity,
Physics Reports. {\bf 509},  167-321 (2011).

\bibitem{Sotiriou}
$f(R)$ Theories of Gravity,
Rev. Mod. Phys. {\bf 82}, 451-497 (2010).

\bibitem{Vas} 
S.D. Odintsov, V.K. Oikonomou, I. Giannakoudi, F.P. Fronimos, E.C. Lymperiadou,
Recent Advances on Inflation,
Symmetry {\bf 15}, 1701  (2023).

\bibitem{capozziello:overview}
S. Capozziello, F. Bajardi, Non-Local Gravity Cosmology: an Overview, Int. J. Mod. Phys. D (2022).

\bibitem{modesto:super}
L. Modesto, Super-renormalizable gravity, \textit{The Thirteenth Marcel Grossmann Meeting: On Recent Developments in Theoretical and Experimental General Relativity, Astrophysics and Relativistic Field Theories}, 1128--1130 (2015).

\bibitem{modesto:super-quantum}
L. Modesto, Super-renormalizable quantum gravity, Phys. Rev. D \textbf{86}, 044005 (2012).

\bibitem{briscese:inflation}
F. Briscese, A. Marciano, L. Modesto, E. N. Saridakis, Inflation in (super-) renormalizable gravity, Phys. Rev. D \textbf{87}, 083507 (2013).

\bibitem{biswas:bouncing}
T. Biswas, A. Mazumdar, W. Siegel, Bouncing Universes in String-inspired Gravity, J. Cosmol. Astropart. Phys. \textbf{03}, 009 (2006).

\bibitem{modesto:black}
L. Modesto, J. W. Moffat, P. Nicolini, Black holes in an ultraviolet complete quantum gravity, Phys. Lett. B \textbf{695}, 397--400 (2011).


\bibitem{Kuzmin}
Yu. V. Kuzmin,    The convergent nonlocal gravitation, Yad. Fiz. 50, 1630-1635 (1989).

\bibitem{Tomboulis}
E.~T.~Tomboulis,
Superrenormalizable gauge and gravitational theories,
[arXiv:hep-th/9702146 [hep-th]] (1997).


\bibitem{Spyridon}
S.~Talaganis, T.~Biswas and A.~Mazumdar,
Towards understanding the ultraviolet behavior of quantum loops in infinite-derivative theories of gravity,
Class. Quant. Grav. \textbf{32}, 215017 (2015).


\bibitem{Lambiase1}
L.~Buoninfante, A.~Ghoshal, G.~Lambiase and A.~Mazumdar,
Transmutation of nonlocal scale in infinite derivative field theories,
Phys. Rev. D \textbf{99}, 044032 (2019).


\bibitem{Lambiase2}
L.~Buoninfante, G.~Lambiase and M.~Yamaguchi,
Nonlocal generalization of Galilean theories and gravity,
Phys. Rev. D \textbf{100}, 026019 (2019).

\bibitem{biswas:towards}
T. Biswas, E. Gerwick, T. Koivisto, A. Mazumdar, Towards singularity-and ghost-free theories of gravity, Phys. Rev. Lett. \textbf{108}, 031101 (2012).

\bibitem{ModestoTsu}
L.~Modesto and S.~Tsujikawa,
Non-local massive gravity,
Phys. Lett. B \textbf{727} (2013), 48-56

\bibitem{Jaccardgla}
M.~Jaccard, M.~Maggiore and E.~Mitsou,
Nonlocal theory of massive gravity,
Phys. Rev. D \textbf{88}, 044033 (2013).


\bibitem{deser:nonlocal1}
S. Deser, R. P. Woodard, Nonlocal Cosmology, Phys. Rev. Lett. \textbf{99}, 111301 (2007).

\bibitem{deser:nonlocal2}
S. Deser, R. P. Woodard, Nonlocal Cosmology II -- Cosmic acceleration without fine tuning or dark energy, J. Cosmol. Astropart. Phys. \textbf{06}, 034 (2019).

\bibitem{arkani:non-local}
N. Arkani-Hamed, S. Dimopoulos, G. Dvali, G. Gabadadze, Non-Local Modification of Gravity and the Cosmological Constant Problem, arXiv preprint hep-th/0209227 (2002).

\bibitem{barvinsky:nonlocal}
A. O. Barvinsky, Nonlocal action for long-distance modifications of gravity theory, Phys. Lett. B \textbf{572}, 109--116 (2003).

\bibitem{nojiri:screening}
S. Nojiri, S. D. Odintsov, M. Sasaki, Y. Zhang, Screening of cosmological constant in non-local gravity, Phys. Lett. B \textbf{696}, 278--282 (2011).

\bibitem{odintsov1}
S.~Nojiri, S.~D.~Odintsov and V.~K.~Oikonomou,
Ghost-free non-local $F(R)$ Gravity Cosmology,
Phys. Dark Univ. \textbf{28} (2020), 100541 (2020).

\bibitem{odintsov2}
E.~Elizalde, S.~D.~Odintsov, E.~O.~Pozdeeva and S.~Y.~Vernov,
De Sitter and power-law solutions in non-local Gauss\textendash{}Bonnet gravity,
Int. J. Geom. Meth. Mod. Phys. \textbf{15}, 1850188 (2018).

\bibitem{odintsov3}
K.~Bamba, S.~Nojiri, S.~D.~Odintsov and M.~Sasaki,
Screening of cosmological constant for De Sitter Universe in non-local gravity, phantom-divide crossing and finite-time future singularities,
Gen. Rel. Grav. \textbf{44}, 1321-1356 (2012).

\bibitem{Lev}
F.M. Lev, 
Solving Particle–Antiparticle and Cosmological Constant Problems, Axioms, \textbf{13}, 138 (2024).

\bibitem{rocco}
S.~Capozziello and R.~D'Agostino,
Reconstructing the distortion function of non-local cosmology: A model-independent approach,
Phys. Dark Univ. \textbf{42}, 101346 (2023).

\bibitem{biswas:stable}
T. Biswas, A. S. Koshelev, A. Mazumdar, S. Y. Vernov, Stable bounce and inflation in non-local higher derivative cosmology, J. Cosmol. Astropart. Phys. \textbf{08}, 024 (2012).

\bibitem{Novikov1}
E.~A.~Novikov,
Ultralight gravitons with tiny electric dipole moment are seeping from the vacuum,
Mod. Phys. Lett. A \textbf{31}, 1650092 (2016).


\bibitem{Novikov2}
E.~A.~Novikov,
Quantum Modification of General Relativity,
Electron. J. Theor. Phys. \textbf{13}, 79-90 (2016).

\bibitem{Novikov3}
 E.  A. Novikov, Big Bang?, Journal of High Energy Physics, Gravitation and Cosmology \textbf{9 },  964 (2023). 
  

\bibitem{capriolo1}
S.~Capozziello, M.~Capriolo and S.~Nojiri,
Considerations on gravitational waves in higher-order local and non-local gravity,
Phys. Lett. B \textbf{810}, 135821 (2020).

\bibitem{capriolo2}
S.~Capozziello, M.~Capriolo and G.~Lambiase,
The energy\textendash{}momentum complex in non-local gravity,'
Int. J. Geom. Meth. Mod. Phys. \textbf{20}, 2350177 (2023).

\bibitem{maggioreGW}
E.~Belgacem, Y.~Dirian, A.~Finke, S.~Foffa and M.~Maggiore,
Nonlocal gravity and gravitational-wave observations,
JCAP \textbf{11} (2019), 022.

\bibitem{Wood}
C. Deffayet, R. P. Woodard, 
The Price of Abandoning Dark Matter Is Nonlocality, 
arXiv:2402.11716 [gr-qc] (2024).

\bibitem{kostas}
K.~F.~Dialektopoulos, D.~Borka, S.~Capozziello, V.~Borka Jovanovi\'c and P.~Jovanovi\'c,
Constraining nonlocal gravity by S2 star orbits,
Phys. Rev. D \textbf{99}, 044053, (2019).

\bibitem{filippo1}
F.~Bouch\`e, S.~Capozziello, V.~Salzano and K.~Umetsu,
Testing non-local gravity by clusters of galaxies,
Eur. Phys. J. C \textbf{82} (2022) no.7, 652, (2022).

\bibitem{filippo2}
F.~Bouch\`e, S.~Capozziello and V.~Salzano,
Addressing Cosmological Tensions by Non-Local Gravity,
Universe \textbf{9}, 27. (2023).



\bibitem{maggiore:phantom}
M. Maggiore, Phantom dark energy from nonlocal infrared modifications of general relativity, Phys. Rev. D \textbf{89} (2014).

\bibitem{foffa:cosmological}
S. Foffa, M. Maggiore, E. Mitsou, Cosmological dynamics and dark energy from nonlocal infrared modifications of gravity, Int. J. Mod. Phys. A \textbf{29}, 1450116 (2014).

\bibitem{maggiore:nonlocal}
M. Maggiore, M. Mancarella, Nonlocal gravity and dark energy, Phys. Rev. D \textbf{90}, 023005 (2014).

\bibitem{jaccard:nonlocal}
M. Jaccard, M. Maggiore, E. Mitsou, Nonlocal theory of massive gravity, Phys. Rev. D \textbf{88}, 044033 (2013).

\bibitem{biswas2}
T.~Biswas, A.~Conroy, A.~S.~Koshelev and A.~Mazumdar,
Generalized ghost-free quadratic curvature gravity,
Class. Quant. Grav. \textbf{31}, 015022 (2014).
[erratum: Class. Quant. Grav. \textbf{31}, 159501 (2014)]

\bibitem{koshelev:bouncing}
A. S. Koshelev, S. Y. Vernov, On Bouncing Solutions in Non-local Gravity, Phys. Part. Nucl. \textbf{43}, 666--668 (2012).

\bibitem{vanPutten3}
M.~H.~P.~M.~van Putten,
The Fast and Furious in JWST high-z galaxies,
Phys. Dark Univ. \textbf{43}, 101417 (2024).

\bibitem{JWST}
D.J. Eisenstein et al.,
Overview of the JWST Advanced Deep Extragalactic Survey (JADES),
[arXiv:2306.02465 [astro-ph.CO]] (2023).

\bibitem{DESI}
A.G. Adame et al.[DESI], DESI2024VI:Cosmological Constraints from the Measurements of Baryon Acoustic Oscillations,
 [arXiv:2404.03002 [astro-ph.CO]].
 
\bibitem{biswas:propagator}
T. Biswas, T. Koivisto, A. Mazumdar, Nonlocal theories of gravity: the flat space propagator, \textit{Proceedings of the Barcelona Postgrad Encounters on Fundamental Physics}, 13--24 (2013).

\bibitem{Belgacem}
E.~Belgacem, A.~Finke, A.~Frassino and M.~Maggiore,
Testing nonlocal gravity with Lunar Laser Ranging,\\
JCAP \textbf{02} (2019), 035.


\bibitem{Koshelev4}
A. S. Koshelev, L. Modesto, L. Rachwal, A. A. Starobinsky,
Occurrence of exact $R^2$ inflation in non-local UV-complete gravity, 
JHEP {\bf 11},  1 (2016).

\bibitem{Koshelev2}
A. S. Koshelev, K. Sravan Kumar, A. A. Starobinsky,
Generalized non-local $R^2$-like inflation,
JHEP {\bf 07},  146 (2023).

\bibitem{Koshelev1}
A. S. Koshelev, K. Sravan Kumar, A. A. Starobinsky,
Non-Gaussianities in generalized non-local $R^2$-like inflation,
JHEP {\bf 07},  094 (2023).

\bibitem{Koshelev3}
 A. S. Koshelev, K. Sravan Kumar, A. Mazumdar, A. A. Starobinsky,
 Non-Gaussianities and tensor-to-scalar ratio in non-local $R^2$-like inflation, 
JHEP {\bf 06}, 152 (2020).


\bibitem{VasA}
V. K. Oikonomou, 
Rescaled Einstein-Hilbert Gravity from f(R) Gravity: Inflation, Dark Energy and the Swampland Criteria, 
Phys. Rev. D {\bf 103}, 124028 (2021).

\bibitem{VasB}
S.D. Odintsov, V.K. Oikonomou, and F.P. Fronimos, 
Inflationary Dynamics and Swampland Criteria for Modified Gauss-Bonnet Gravity Compatible with GW170817,
Phys. Rev. D {\bf 107}, 084007 (2023).

\bibitem{VasC}
V.K. Oikonomou, I. Giannakoudi, A. Gitsis, K. R. Revis, 
Rescaled Einstein-Hilbert Gravity: Inflation and the Swampland Criteria, 
Int. J. of Mod. Phys.  D {\bf 31},  2250001 (2022).

 \end{thebibliography}
\end{document}